\documentclass[preprints,article,accept,pdftex,moreauthors]{Definitions/mdpi} 

\usepackage{graphicx}
\usepackage{multicol}
\usepackage{dcolumn}
\usepackage{bm}
\usepackage{amsmath}
\usepackage{amssymb}
\usepackage{float}
\usepackage{lipsum}
\usepackage{color}
\usepackage{textcomp}
\usepackage{latexsym}
\usepackage{mathtools}
\usepackage{url}
\usepackage{acro}
\usepackage{adjustbox}
\usepackage{array}

\DeclareAcronym{NANOGrav}{
  short = NANOGrav ,
  long = North American Nanohertz Observatory for Gravitational Waves ,
  short-plural =  ,
}
\DeclareAcronym{LVK}{
  short = LVK ,
  long = {Advanced LIGO, Virgo and KAGRA} ,
  short-plural = ,
}
\DeclareAcronym{GW}{
  short = GW ,
  long = gravitational wave ,
  short-plural = s ,
}
\DeclareAcronym{SIGW}{
  short = SIGW ,
  long = scalar induced gravitational wave ,
  short-plural = s ,
}
\DeclareAcronym{SGWB}{
  short = SGWB ,
  long = stochastic gravitational-wave background ,
  short-plural = s ,
}
\DeclareAcronym{CBC}{
  short = CBC ,
  long = compact binary coalescence ,
  short-plural = s ,
}
\DeclareAcronym{BH}{
  short = BH ,
  long = black hole ,
  short-plural = s ,
}
\DeclareAcronym{BBH}{
  short = BBH ,
  long = binary black hole ,
  short-plural = s ,
}
\DeclareAcronym{PBH}{
  short = PBH ,
  long = primordial black hole ,
  short-plural = s ,
}
\DeclareAcronym{SNR}{
  short = SNR ,
  long = signal-to-noise ratio ,
  short-plural = s ,
}
\DeclareAcronym{IMRPPv2}{
  short = ,
  long = {\normalsize IMRP}{\footnotesize HENOM}{\normalsize P}v2 ,
  short-plural = ,
}
\DeclareAcronym{PTA}{
  short = PTA ,
  long = pulsar timing array ,
  short-plural = s ,
}
\DeclareAcronym{SFR}{
  short = SFR ,
  long = star formation rate ,
  short-plural =  ,
}
\DeclareAcronym{FRW}{
  short = FRW ,
  long = Friedman-Robertson-Walker ,
  short-plural =  ,
}
\DeclareAcronym{IMR}{
  short = IMR ,
  long = inspiral-merger-ringdown ,
  short-plural =  ,
}
\DeclareAcronym{LISA}{
	short = LISA ,
	long  = Laser Interferometer Space Antenna,
  short-plural =  ,
}
\DeclareAcronym{ET}{
	short = ET ,
	long  = Einstein Telescope,
  short-plural =  ,
}
\DeclareAcronym{SKA}{
	short = SKA ,
	long  = Square Kilometer Array,
  short-plural =  ,
}
\DeclareAcronym{FAST}{
	short = FAST ,
	long  = Five-hundred-meter Aperture Spherical radio Telescope,
  short-plural =  ,
}
\DeclareAcronym{CE}{
	short = CE ,
	long  = Cosmic Explorer,
  short-plural =  ,
}
\DeclareAcronym{BBO}{
	short = BBO ,
	long  = Big Bang Observer,
  short-plural =  ,
}
\DeclareAcronym{DECIGO}{
	short = DECIGO ,
	long  = Deci-hertz Interferometer Gravitational wave Observatory,
  short-plural =  ,
}
\DeclareAcronym{ABH}{
	short = ABH ,
	long  = astrophysical black hole,
  short-plural = s ,
}
\DeclareAcronym{SMBH}{
	short = SMBH ,
	long  = supermassive black hole,
  short-plural = s ,
}
\DeclareAcronym{GR}{
	short = GR ,
	long  = general relativity,
  short-plural =  ,
}
\DeclareAcronym{CMB}{
	short = CMB ,
	long  = cosmic microwave background ,
  short-plural =  ,
}
\DeclareAcronym{LSS}{
	short = LSS ,
	long  = large-scale structure ,
  short-plural = s ,
}
\DeclareAcronym{BBN}{
	short = BBN ,
	long  = big bang nucleosynthesis ,
  short-plural =  ,
}
\DeclareAcronym{SSE}{
	short = SSE ,
	long  = solar system ephemeris
}
\DeclareAcronym{JPL}{
	short = JPL ,
	long  = Jet Propulsion Laboratory
}
\DeclareAcronym{TOA}{
	short = TOA ,
	long  = time of arrival,
  short-plural = s ,
}
\DeclareAcronym{ORF}{
	short = ORF ,
	long  = overlap reduction function,
  short-plural = s ,
}
\DeclareAcronym{PSD}{
	short = PSD ,
	long  = power spectral density,
  short-plural = s ,
}
\firstpage{1} 
\pubvolume{1}
\issuenum{1}
\articlenumber{0}
\pubyear{2023}
\copyrightyear{2023}
\externaleditor{Academic Editor: } 
\datereceived{18 February 2023} 
\daterevised{20 March 2023} % Comment out if no revised date
\dateaccepted{ } 
\datepublished{ } 
\hreflink{https://doi.org/} % If needed use \linebreak
\Title{Bayesian Implications for the Primordial Black Holes from NANOGrav's Pulsar-Timing Data Using the Scalar-Induced Gravitational Waves}

\TitleCitation{Bayesian Implications for the Primordial Black Holes from NANOGrav's Pulsar-Timing Data Using the Scalar-Induced Gravitational Waves}
\def\({\left(}
\def\){\right)}
\def\[{\left[}
\def\]{\right]}

\newcommand{\be}{{\begin{eqnarray}}}
\newcommand{\ee}{{\end{eqnarray}}}

\newcommand{\overbar}[1]{\mkern 1.5mu\overline{\mkern-1.5mu#1\mkern-1.5mu}\mkern 1.5mu}

\Author{{Zhi-Chao Zhao $^{1,\dagger}$\orcidA{}} 
 and Sai Wang $^{2,3,}$*$^{,\dagger}$\orcidB{}}

\AuthorNames{Zhi-Chao Zhao and Sai Wang}

\AuthorCitation{Zhao, Z.-C.; Wang, S.}
\address{%
$^{1}$ \quad Department of Applied Physics, College of Science, China Agricultural University, Qinghua East Road, Beijing 100083, China; {zhaozc@cau.edu.cn} 
\\
$^{2}$ \quad Theoretical Physics Division, Institute of High Energy Physics, Chinese Academy of Sciences, \linebreak Beijing 100049, China\\
$^{3}$ \quad School of Physical Sciences, University of Chinese Academy of Sciences, Beijing 100049, China}

\corres{Correspondence: wangsai@ihep.ac.cn}

\firstnote{These authors contributed equally to this work.} 
\abstract{Assuming that the common-spectrum process in the NANOGrav 12.5-year dataset has an origin of scalar-induced gravitational waves, we study the enhancement of primordial curvature perturbations and the mass function of primordial black holes, by performing the Bayesian parameter inference for the first time. We obtain lower limits on the spectral amplitude, i.e., $\mathcal{A}\gtrsim10^{-2}$ at 95\% confidence level, when assuming the power spectrum of primordial curvature perturbations to follow a log-normal distribution function with width $\sigma$. In the case of $\sigma\rightarrow0$, we find that the primordial black holes with $2\times10^{-4}-10^{-2}$ solar mass are allowed to compose at least a fraction $10^{-6}$ of dark matter. Such a mass range is shifted to more massive regimes for larger values of $\sigma$, e.g., to a regime of $4\times10^{-3}-0.2$ solar mass in the case of $\sigma=1$. 
We expect the planned gravitational-wave experiments to have their best sensitivity to $\mathcal{A}$ in the range of $10^{-4}$ to $10^{-7}$, depending on the experimental setups. With this level of sensitivity, we can search for primordial black holes throughout the entire parameter space, especially in the mass range of $10^{-16}$ to $10^{-11}$ solar masses, where they could account for all dark matter. 
In addition, the importance of multi-band detector networks is emphasized to accomplish our theoretical expectation. }


\begin{document}
\crefname{figure}{Fig.}{Figs.}
\Crefname{figure}{Fig.}{Figs.}

\section{Introduction}\label{sec:intr}

It is well-known that the \acp{PBH} could be produced by gravitational collapses of the enhanced primordial curvature perturbations on small scales~\cite{Hawking:1971ei}. 
The mass function of \acp{PBH} is determined by the power spectrum of primordial curvature perturbations \cite{Khlopov:2008qy,Belotsky:2014kca,Franciolini:2022tfm}. 
To produce a sizeable quantity of \acp{PBH}, the scalar spectral amplitude should conquer a critical magnitude $\sim$$0.1$~\cite{Musco:2020jjb}, which is about eight orders of magnitudes larger than the spectral amplitude on large scales. 
On the other hand, the dynamics of cosmic inflation~\cite{Guth:1980zm,Linde:1981mu,Starobinsky:1980te} is one of the greatest puzzles of cosmology to be uncovered. 
In particular, the small-scale primordial curvature perturbations were generated by quantum fluctuations at the late-time stage of inflation. 
Their power spectrum encodes characteristic information about the late-time dynamics of inflation, which could be different from that of the early-time one measured precisely with experiments on \ac{CMB}~\cite{WMAP:2012nax,Planck:2018vyg} and \ac{LSS}~\cite{BOSS:2016wmc,DES:2021wwk}.  
Therefore, we can gain more information about cosmic inflation by  investigating the \acp{PBH}.

The scenario of \acp{PBH} can also be a reasonable candidate of dark matter, or compose a part of the latter~\cite{Carr:1974nx,Chapline:1975ojl}. The mass function of \acp{PBH} is defined as a ratio between the energy density fractions of \acp{PBH} and dark matter. 
Extensive studies have shown observational constraints on the mass function of \acp{PBH} (e.g., see reviews in Refs.~\cite{Carr:2020gox,Escriva:2022duf}). 
In addition, the scenario of \acp{PBH} was suggested to interpret the formation mechanism of \acp{SMBH}~\cite{Kohri:2022wzp} and the events of \aclp{CBC}~\cite{Sasaki:2016jop,Raidal:2017mfl,Ali-Haimoud:2017rtz,Chen:2018czv,Nishikawa:2017chy,Bird:2016dcv,Wang:2021iwp,Belotsky:2018wph}, which emitted the \acp{GW} observed by the \acl{LVK} detectors~\cite{Harry:2010zz,VIRGO:2014yos,Somiya:2011np}.

Accompanied by the formation process of \acp{PBH}, the \acp{SIGW}~\cite{Mollerach:2003nq,Ananda:2006af,Baumann:2007zm,Assadullahi:2009jc} were inevitably produced by the enhanced primordial curvature perturbations. 
Specifically, they were produced through nonlinear couplings of tensor-scalar-scalar modes in the very early cosmos. 
Given the power spectrum of primordial curvature perturbations, a semi-analytic formula was provided to compute the energy density fraction spectrum of \acp{SIGW}~\cite{Kohri:2018awv,Espinosa:2018eve}. 
It has been widely used to investigate the mass function of \acp{PBH} in the literature.

In this work, we will investigate the mass function of \acp{PBH}, as well as the power spectrum of primordial curvature perturbations, by analyzing a 12.5-year dataset from \ac{NANOGrav}\endnote{{\url{http://data.nanograv.org}, version v3} 
}~\cite{Arzoumanian:2020vkk}. 
Strong evidence of a stochastic common-spectrum process was reported by \ac{NANOGrav} Collaboration. 
%A Bayes factor was shown to exceed $10^4$ in favor of the common-spectrum process versus independent red-noise processes. 
If this signal has an origin of \acp{SIGW}, we can straightforwardly obtain constraints on the power spectrum of primordial curvature perturbations. 
Since the latter can collapse to form the \acp{PBH} due to gravity, we therefore obtain corresponding constraints on the mass function of \acp{PBH}. 
In fact, the scenario of \acp{PBH} with different masses has been suggested to interpret such a signal (see, e.g., Refs.~\cite{DeLuca:2020agl,Vaskonen:2020lbd,Kohri:2020qqd,Domenech:2020ers,Atal:2020yic,Yi:2022ymw}). 
However, there is still lack of systematic Bayesian parameter inferences, which will be performed for the first time in this paper. 

%\endnote{The authors of Ref.~\cite{Chen:2019xse} studied the mass function of \acp{PBH} via searching for the \acp{SIGW} in the \ac{NANOGrav} 11-year dataset. At last stage of preparation of our paper, the authors of Ref.~\cite{Yi:2022ymw} analyzed the \ac{NANOGrav} 12.5-year dataset to study the primordial curvature spectrum with different templates of scalar induced gravitational waves.}

The rest of the paper is arranged as follows. We briefly review the theory of \acp{SIGW} in Section~\ref{sec:sigw}. By performing Bayesian analysis, we obtain the \ac{NANOGrav} constraints on the enhancement of primordial curvature perturbations in Section~\ref{sec:cpcp}. Correspondingly, we demonstrate the physical implications for the mass function of \acp{PBH} in Section~\ref{sec:cpbh}. We forecast the sensitivity curves of ongoing and planned gravitational-wave detectors in Section~\ref{sec:pros}. The conclusion and discussion are shown in Section~\ref{sec:conc}.

\section{Theory of Scalar-Induced Gravitational Waves}\label{sec:sigw}

As previously shown \cite{Kohri:2018awv,Espinosa:2018eve}, there is a lengthy and tedious derivation process to obtain a semi-analytic formula for the energy density fraction spectrum of \acp{SIGW} in the early cosmos. 
Therefore, we summarize some theoretical results that would be essential to our data analysis. 
The energy density fraction spectrum of \acp{GW} is defined as \linebreak $\Omega_{\mathrm{GW}}(k)=\rho_{\mathrm{GW}}(k)/\rho_{c}$, where $k$ denotes the \ac{GW} wavenumber, $\rho_{c}$ is the critical energy density of the cosmos, and $\int\rho_{\mathrm{GW}}(k)d\ln k$ denotes the total energy density of \acp{GW}. 
The \acp{SIGW} in the subhorizon have the energy density spectrum as $\rho_{\mathrm{GW}}\sim\langle \overbar{h_{ij,l} h_{ij,l}} \rangle$~\cite{Maggiore:2007ulw}, where the overbar stands for the oscillation average and the angle brackets defines the power spectrum. 
The strain of \acp{SIGW}, as the second-order tensor perturbations produced by the linear scalar perturbations, can be roughly expressed as $h\sim \phi^2$, where $\phi$ denotes the linear scalar perturbation. 
If considering the Gaussian scalar perturbations, we obtain $\rho_{\mathrm{GW}}\sim\langle \phi^{4} \rangle \sim \langle \phi^{2} \rangle \langle \phi^{2} \rangle$, where $\langle \phi^2 \rangle$ is determined by the initial value and transfer function of $\phi$. 
Therefore, following Equation~(14) of Ref.~\cite{Kohri:2018awv}, the energy density fraction spectrum of \acp{SIGW} in the radiation-dominated cosmos is given as   
\begin{eqnarray}\label{eq:ogw}
    \Omega_{\mathrm{GW}}(k)
    =
    \int_0^{\infty} d v
    \int_{\lvert 1-v \rvert} ^{\lvert 1+v \rvert} d u
    \ F(v,u)\mathcal{P}_{\mathcal{R}}(vk) \mathcal{P}_{\mathcal{R}}(uk)\ , \label{eq:eds}
\end{eqnarray}
where $\mathcal{P}_{\mathcal{R}}(uk)$ (and $\mathcal{P}_{\mathcal{R}}(vk)$) is power spectrum of primordial curvature perturbations at wavenumber $uk$ (and $vk$) with $u$ (and $v$) being a dimensionless variable, and $F(v,u)$ is a functional of the scalar transfer function describing the post-inflationary evolution of $\phi$ and can be simplified as  

\begin{eqnarray}
    F(v,u)
    &=& \frac{3}{2^{10} v^{8}u^{8}}\left[4v^2-(v^2-u^2+1)^{2}\right]^{2} \left(v^2+u^2-3\right)^{2}\nonumber\\
    &&\ \times \bigg\{\left[\left(v^{2}+u^{2}-3\right)\ln\left(\left|\frac{3-(v+u)^{2}}{3-(v-u)^{2}}\right|\right)-4vu\right]^{2}\nonumber\\
    &&\ +\pi^2\left(v^{2}+u^{2}-3\right)^{2}\Theta\left(v+u-\sqrt{3}\right)\bigg\}\ .\label{eq:fuv}
\end{eqnarray} 

Due to a lack of direct measurements, the nature of primordial curvature perturbations on small scales remains unknown, compared with that on the largest scales measured by the cosmic microwave background. 
For computational simplicity, we take two concrete expressions of $\mathcal{P}_{\mathcal{R}}(k)$, which are frequently adopted in the literature (e.g., Refs.~\cite{Wang:2019kaf,Bugaev:2009zh,Kohri:2018awv,Adshead:2021hnm,Ferrante:2022mui}), as~follows  
\begin{eqnarray}
    &&\mathcal{P}_{\mathcal{R}}(k)
    = \mathcal{A}\ \delta\left(\ln{\frac{k}{k_\ast}}\right)\label{eq:delta}\ ,\\
     &&\mathcal{P}_{\mathcal{R}}(k)= \frac{\mathcal{A}}{\sqrt{2\pi}\sigma}\exp{\left(-\frac{\ln^2\left( k/k_\ast \right)}{2\sigma^2}\right)}\ , \label{eq:lognormal}
\end{eqnarray}
where $\mathcal{A}$ is an amplitude, 
$\sigma$ is a standard variance, 
and $k_\ast$ is a pivot wavenumber. 
We can relate the wavenumber with the frequency, i.e., $k=2\pi f$ and $k_\ast=2\pi f_\ast$. 
During the process of parameter inferences, both $\mathcal{A}$ and $f_\ast$ are independent parameters. 
However, we will not let $\sigma$ be independent, even though different values of $\sigma$ would alter our results. 
In fact, it is challenging to explore the full parameter space of $\sigma$ based on the current dataset. 
Therefore, we let $\sigma=1$ for simplicity in this work. 
We can test the robustness of our conclusions by comparing the results obtained when taking two different values of $\sigma$. 
In particular, we will compare the results from Equations~(\ref{eq:delta}) and~(\ref{eq:lognormal}). 
We leave this point to be demonstrated further in the following sections.

\begin{figure}[H]
    
    \includegraphics[width=1.0\columnwidth]{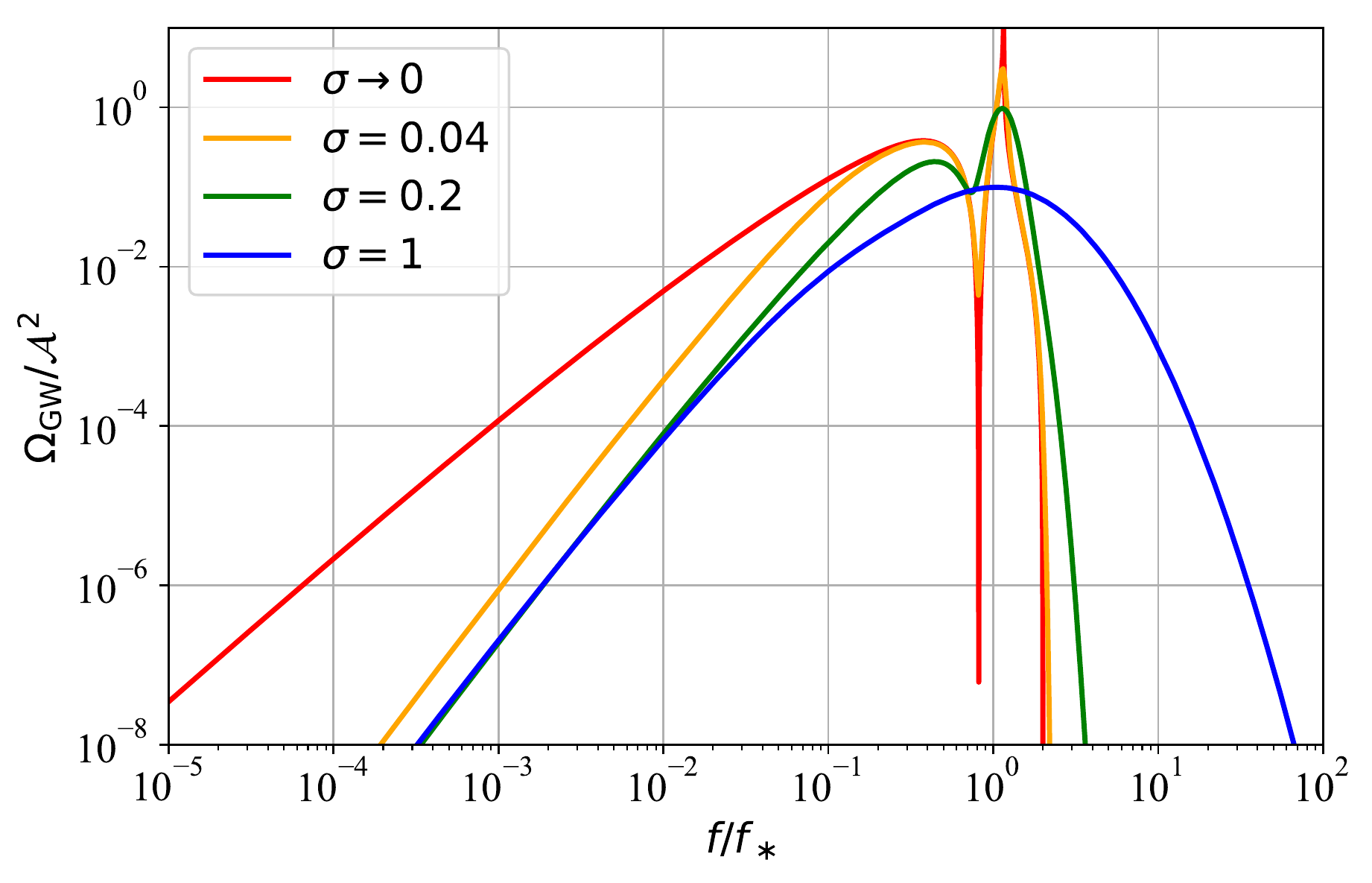}
    \caption{Energy density fraction spectrum of SIGWs normalized with ${\mathcal{A}}^{2}$. We adopt $\sigma\rightarrow0$ to label the power spectrum of primordial curvature perturbations in Equation~(\ref{eq:delta}). Different choices of $\sigma$ in Equation~(\ref{eq:lognormal}) are plotted for comparison. 
    }
    \label{fig:fig2}
\end{figure}

In Figure~\ref{fig:fig2}, we show the energy density fraction spectrum of \acp{SIGW}, which is normalized with $\mathcal{A}^{2}$. 
For comparison, we show the results when taking different versions of $\sigma$.  %please confirm in case of changed meaning
Equation~(\ref{eq:delta}) can be viewed as a limit of Equation~(\ref{eq:lognormal}) when $\sigma\rightarrow0$. 
In the regime of $f\ll f_\ast$, the choice of $\sigma\rightarrow0$ displays the highest spectral amplitude than non-vanishing choices of $\sigma$. 
This implies that in such a case, we can obtain the best sensitivity on measurements of $\mathcal{A}$ for a given \ac{GW} experiment. Correspondingly, we can obtain the strictest constraints on the mass function of \acp{PBH}. 
We will demonstrate these points in the following sections.

Once the early-universe quantity $\Omega_{\mathrm{GW}}(k)$ is given in Equation~(\ref{eq:ogw}), we have the energy density fraction spectrum of \acp{SIGW} in the present cosmos to be~\cite{Wang:2019kaf}
\begin{equation}\label{eq:ogw0}
    \Omega_{\mathrm{GW},0}(f) = \Omega_{\mathrm{r},0}\left[ \frac{g_{\ast,\rho}(T)}{g_{\ast,\rho}(T_{\rm eq})} \right] \left[ \frac{g_{\ast,s}(T_{\rm eq})}{g_{\ast,s}(T)} \right]^{\frac{4}{3}}  \Omega_{\mathrm{GW}}(k)\ ,
\end{equation}
where $\Omega_{\mathrm{r},0}$ is the present energy density fraction of radiations, and the subscript $_{\rm eq}$ stands for cosmological quantities at the epoch of matter-radiation equality. 
Here, we have taken into account contributions from the effective relativistic degrees of the cosmos, i.e., $g_{\ast,\rho}$ and $g_{\ast,s}$, as functions of $f$ by interpolating their tabulated data in terms of cosmic temperature $T$ in Ref.~\cite{Saikawa:2018rcs}\endnote{{\url{https://member.ipmu.jp/satoshi.shirai/EOS2018.php}, accessed on 1 January 2019} 
} as well as by considering a relation between $T$ and $f$ in Wang et al. ~\cite{Wang:2019kaf}, i.e., 
\begin{equation}
    \frac{f}{\mathrm{nHz}} = 26.5 \left(\frac{T}{\mathrm{GeV}} \right)\left[\frac{g_{\ast,\rho}(T)}{106.75}\right]^{\frac{1}{2}} \left[\frac{g_{\ast,s}(T)}{106.75}\right]^{-\frac{1}{3}} \ .\label{eq:fT}
\end{equation} 
 %is  
%\begin{equation}
%    $\Omega_{\mathrm{r},0} = 4.15\times10^{-5} h^{-2}$%\ ,
%\end{equation}

Throughout this paper, we use the measured value of cosmological parameters in the data release 2018 of \emph{{Planck} 
}satellite~\cite{Planck:2018vyg}. 
The publicly available \texttt{Astropy}~\cite{Astropy:2013muo,Astropy:2018wqo,Astropy:2022ucr}  {software} \endnote{{\url{https://www.astropy.org/}, version 5.1}}  
 is adopted to evaluate all cosmological quantities. 
Please note that Equation~(\ref{eq:ogw0}) is one of the leading formulas that will be used during Bayesian analysis in the next section.

\section{NANOGrav Constraints on Primordial Curvature Perturbations}\label{sec:cpcp}

When a \ac{PTA} experiment measures the stochastic gravitaitonal wave background, the timing residual cross-power spectral density is given by \linebreak $S_{ab}(f)=\Gamma_{ab}h_{c}^{2}/(12\pi^{2}f^{3})$, which can be shown by combining Equations~(1) and (2) in Arzoumanian et al. ~\cite{Arzoumanian:2018saf}. 
Here, $\Gamma_{ab}$ stands for the \ac{ORF} that describes the correlation between two pulsars $a$ and $b$ as a function of line-of-sight separation angle between them. 
Following Equation~(17) in Maggiore~\cite{Maggiore:1999vm}, $h_{c}(f)$ is the characteristic strain defined as $h_{c}^{2}=3H_{0}^{2}\Omega_{\mathrm{GW},0}/(2\pi^{2}f^{2})$, where $H_0=100h_{0}\ \mathrm{km}\ \mathrm{s}^{-1}\mathrm{Mpc}^{-1}$ is the Hubble constant with $h_{0}$ being the reduced Hubble constant. 
Therefore, the timing residual cross-power spectral density becomes~\cite{Arzoumanian:2018saf} 
\begin{equation}\label{Sab}
    S_{ab} = \Gamma_{ab} \frac{H_0^2 f_{\mathrm{yr}}^{-5} }{8 \pi^4}  \left(\frac{f}{f_{\mathrm{yr}}}\right)^{-5} \Omega_{\mathrm{GW},0}(f)\ ,
\end{equation}
where $f_{\mathrm{yr}}$ is a pivot frequency corresponding to a duration time of 1 year, and the formula for $\Omega_{\mathrm{GW},0}(f)$ is shown in Equation~(\ref{eq:ogw0}).   
For an isotropic background of \acp{GW}, e.g., the \acp{SIGW} considered in this work, we take \ac{ORF} to be the Hellings and Downs coefficients~\cite{Hellings:1983fr}.

%{\color{red}
The timing residual data for a single pulsar is decomposed in its individual constituents, i.e.,~\cite{Arzoumanian:2015liz}
\begin{equation}
\delta\bm{t} = M \bm{\epsilon} + F \bm{a} + U\bm{j} + \bm{n}\ . 
\end{equation}

The term $M\bm{\epsilon}$ stands for the inaccuracies in the subtraction of timing model, where $M$ denotes the timing model design matrix, and $\bm{\epsilon}$ is a vector describing small offsets for the timing model parameters.
The matrix $M$ is computed through \texttt{libstempo}~\cite{2020ascl.soft02017V}\endnote{{\url{https://vallis.github.io/libstempo/}, version 2.4.5} 
}, which is a python interface for \texttt{TEMPO2}~\cite{Hobbs:2006cd,Edwards:2006zg}\endnote{{\url{https://bitbucket.org/psrsoft/tempo2.git}, version 2021.07.1} 
} timing package. 
We use the latest \ac{JPL} \ac{SSE} model, DE438~\cite{DE438}, in the timing model fits. 
The term $F\bm{a}$ accounts for all low-frequency signals, including the pulsar-intrinsic red noise. The Fourier design matrix $F$ has alternating sine and cosine functions, and $\bm{a}$ is a vector comprised of Fourier coefficients at the integer ($1,2,\ldots,N_{\mathrm{mode}}$) multiples of the harmonic base frequency $1/T_s$, where $T_s$ denotes the span between the minimum and maximum \acp{TOA} in the array~\cite{vanHaasteren:2014qva}. 
%Similar to Ref.~\cite{Arzoumanian:2020vkk}, we use $N_{\mathrm{mode}}=30$ frequency bands for the power-law spectrum of pulsar-intrinsic red noise, while for the common-spectrum process we use the lowest $N_{\mathrm{mode}}=5$ frequency bands and $N_{\mathrm{mode}}=30$ frequency bands, respectively. 
%to alleviate the effect of potential coupling between the white noise and the highest-frequency bands of the red noise process. 
Described by a per-epoch variance (ECORR) for each receiver and backend system~\cite{Arzoumanian:2015liz}, the term $U\bm{j}$ denotes the white noise which is fully correlated for simultaneous observations at different frequencies, but fully uncorrelated in time.
The matrix $U$ maps all \acp{TOA} observed simultaneously at different frequencies to a total \ac{TOA}, and $\bm{j}$ is the per-epoch white noise which is fully correlated across all observing frequencies. 
The term $\bm{n}$ is the timing residual introduced by Gaussian white noise, which is described by the parameters of the \ac{TOA} uncertainties (EFAC) and an additive variance (EQUAD) for each receiver and backend system~\cite{Arzoumanian:2015liz}. 
%}

%Following Ref.~\cite{Arzoumanian:2018saf,Arzoumanian:2020vkk}, w
To estimate the allowed parameter space, we perform Bayesian parameter inferences by analyzing a dataset of 45 pulsars in the 12.5-year data release of \ac{NANOGrav} Collaboration~\cite{Arzoumanian:2020vkk}\endnote{{\url{https://github.com/nanograv/12p5yr_stochastic_analysis}, accessed on 1 January 2022} 
}. 
We list all independent parameters as well as their priors in Table~\ref{tab:priors}. 
%To relieve the burden on numerical computation, the parameters of white noise are fixed to their best-fits reported by the \ac{NANOGrav} Collaboration\endnote{\url{https://github.com/nanograv/12p5yr_stochastic_analysis}}. 
We adopt $N_{\mathrm{mode}}=30$ frequency bands to the power-law spectrum of pulsar-intrinsic red noise and the common-spectrum process. 
In practice, we will use the publicly available \texttt{enterprise}~\cite{enterprise}%\endnote{\url{https://zenodo.org/record/4059815##.YG2yRi2B0y8}} 
\endnote{{\url{https://zenodo.org/record/4059815}, version 3.2.3}
} 
%and \texttt{enterprise\_extension}\endnote{\url{https://github.com/nanograv/enterprise_extensions}} 
to compute the likelihoods and \texttt{PTMCMCSampler}~\cite{justin_ellis_2017_1037579}%\endnote{\url{https://zenodo.org/record/1037579##.YG2yqS2B0y8}} to perform Markov-Chain Monte-Carlo sampling. 
\endnote{{\url{https://zenodo.org/record/1037579}, version 2.0.0} 
} to perform Markov-Chain Monte-Carlo sampling.

\begin{table}[H]

\tablesize{\small}
\caption{{Priors} 
 used in all analyses performed in this paper.}
\label{tab:priors}
\setlength{\cellWidtha}{\textwidth/4-2\tabcolsep-0.60in}
\setlength{\cellWidthb}{\textwidth/4-2\tabcolsep+.30in}
\setlength{\cellWidthc}{\textwidth/4-2\tabcolsep+.10in}
\setlength{\cellWidthd}{\textwidth/4-2\tabcolsep+.20in}
\scalebox{1}[1]{\begin{tabularx}{\textwidth}{>{\PreserveBackslash\centering}m{\cellWidtha}>{\PreserveBackslash\centering}m{\cellWidthb}>{\PreserveBackslash\centering}m{\cellWidthc}>{\PreserveBackslash\centering}m{\cellWidthd}}
\toprule
\textbf{Parameter} & \textbf{Description} & \textbf{Prior} & \textbf{Comment} \\
\midrule
\multicolumn{4}{c}{White Noise} \\
\midrule
$E_{k}$ & EFAC per backend/receiver system & Uniform $[0, 10]$ & \mbox{single-pulsar analysis only} \\
$Q_{k}$ [s] & EQUAD per backend/receiver system & log-Uniform $[-8.5, -5]$ & \mbox{single-pulsar analysis only} \\
$J_{k}$ [s] & ECORR per backend/receiver system & log-Uniform $[-8.5, -5]$ & \mbox{single-pulsar analysis only} \\
\midrule
\multicolumn{4}{c}{Red Noise} \\
\midrule
$A_{\rm red}$ & power-law spectral amplitude & log-Uniform $[-20, -11]$ & one parameter per pulsar  \\
$\gamma_{\rm red}$ & power-law spectral index & Uniform $[0, 7]$ & one parameter per pulsar \\
\midrule
\multicolumn{4}{c}{Primordial curvature perturbations} \\
\midrule
$\log \mathcal{A}$  & spectral amplitude & Uniform $[-3, 0]$ & one parameter for PTA \\
$\log f_\ast$ [Hz] & spectral pivot frequency & Uniform $[-10,0]$ & one parameter for PTA \\
\bottomrule
\end{tabularx}}

\end{table}

\vspace{-3pt}

\begin{figure}[H]
   
   \hspace{-6pt}  \includegraphics[width=1.0\columnwidth]{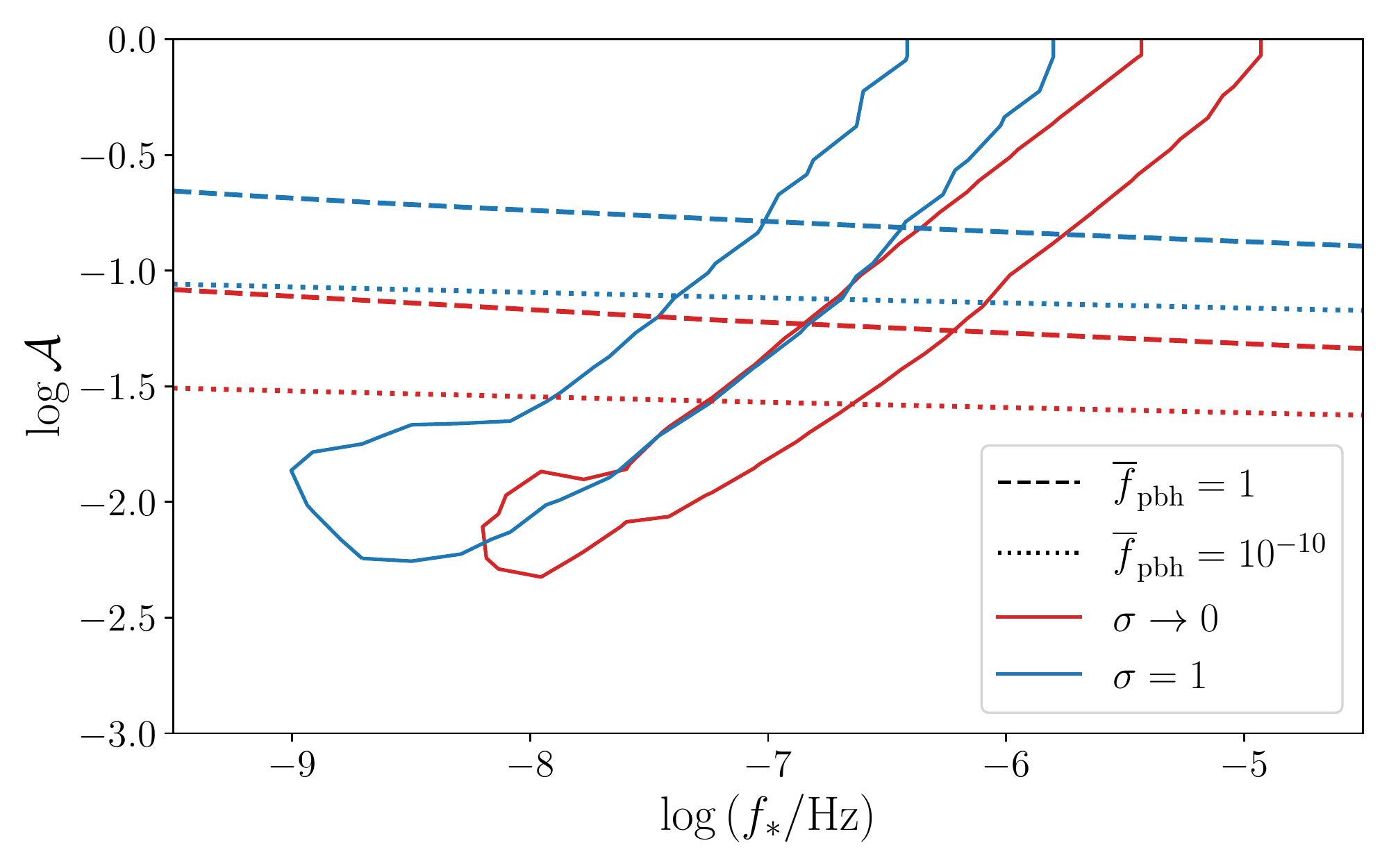}
    \caption{The 95\% CL contour plot of $\mathcal{A}$ and $f_\ast$ inferred from the NANOGrav 12.5yr dataset. Solid curves stand for $\sigma\rightarrow0$ and $\sigma=1$, while other curves denote $\overbar{f}_{\mathrm{pbh}}=1$ and $\overbar{f}_{\mathrm{pbh}}=10^{-10}$. }
    \label{fig:fig1}
\end{figure}

Our results are shown in Figure~\ref{fig:fig1}, which displays the 95\% CL contour plots of $\log\mathcal{A}$ and $\log f_\ast$ for the power spectra of primordial curvature perturbations with $\sigma\rightarrow0$ (red solid curve) and $\sigma=1$ (blue solid curve), respectively. 
We find that $\mathcal{A}$ has lower limits around $10^{-2}$ for the two different choices of $\sigma$. 
%and $f_\ast=(?\pm?)\times10^{?}$ within 68\% confidence interval.  
%In Fig.~\ref{fig:fig3}, we depict a contour of $\mathcal{A}$ and $f_\ast$ for $\mathcal{P}_{\mathcal{R}}(f)$ being log-normal function of $k$ with $\sigma=1$, and obtain $\mathcal{A}>???$.
%and $f_\ast=(?\pm?)\times10^{?}$. 
%By comparing the above two sets of results, we find that the former provides the better sensitivity on measurements of $\mathcal{A}$ than the latter. 
In contrast, the \ac{NANOGrav} dataset prefers different frequency bands. 
For a given value of $\mathcal{A}$, the frequency range in the case of $\sigma\rightarrow0$ is almost always larger than that in the case of $\sigma=1$ by a few times. 
This result is consistent with the expectations of Figure~\ref{fig:fig2}. 
%The frequency band $?<f_\ast/\mathrm{Hz}<?$ is preferred in the case of $\sigma\rightarrow0$ while $?<f_\ast/\mathrm{Hz}<?$ is preferred in the other case.  
However, we find strong positive correlations between $\log\mathcal{A}$ and $\log f_\ast$ in either choice of $\sigma$. 
%However, to make \acp{PBH} to be a significant ingredient of dark matter, we find that the current dataset supports $f_\ast\sim10^{-6?}$ Hz, corresponding to the \ac{PBH} mass range $m \sim 10^{-6?}M_\odot$ that will be discussed in details in the next section. 
Furthermore, in Figure~\ref{fig:fig1}, we also label the spectral amplitudes that produce the \acp{PBH} with total abundance $\overbar{f}_{\mathrm{pbh}}=1$ (dashed curves) and $\overbar{f}_{\mathrm{pbh}}=10^{-10}$ (dotted curves) that will be interpreted in the next section. 

\section{Implications for Primordial Black Holes}\label{sec:cpbh}

We define the mass function of \acp{PBH} as  $f_{\mathrm{pbh}}(m)=\Omega_{\mathrm{dm}}^{-1}d\Omega_{\mathrm{pbh}}/d\ln m$, in which $\Omega_{\mathrm{dm}}$ and $\Omega_{\mathrm{pbh}}$ stand for the present energy density fractions of dark matter and \acp{PBH} with mass $m$, respectively. 
We adopt the concept of critical collapse~\cite{Carr:2016drx,Yokoyama:1998xd} and Press-Schechter formalism~\cite{Press:1973iz}. 
Therefore, following Equation~(9) in Wang et al. ~\cite{Wang:2019kaf}, we have  
\begin{eqnarray}
    f_{\mathrm{pbh}}(m) = \frac{\Omega_{\mathrm{m}}}{\Omega_{\mathrm{dm}}} \int g(T) \tilde{\beta}(m,m_{H}) d\ln m_{H}\ ,
\end{eqnarray}
where $\Omega_{\mathrm{m}}$ is the present energy density fraction of non-relativistic matter.
Here, we denote 
\begin{eqnarray}
    g(T) &=& \frac{g_{\ast,\rho}(T)}{g_{\ast,\rho}(T_{\mathrm{eq}})} \frac{g_{\ast,s}(T_{\mathrm{eq}})}{g_{\ast,s}(T)} \frac{T}{T_{\mathrm{eq}}} \ ,\\
    \tilde{\beta}(m,m_{H}) &=& \frac{\kappa\mu^{\gamma+1}}{\sqrt{2\pi}\gamma \Delta(k)} \mathrm{exp}\left[-\frac{\left(\delta_{c}+\mu\right)^{2}}{2\Delta^{2}(k)}\right] \ ,
\end{eqnarray}
where we have $\mu=[{m}/({\kappa m_{H}})]^{1/\gamma}$ with numerical constants $\kappa=3.3$ and $\gamma=0.36$, and the critical overdensity for gravitational collapse is $\delta_{c}=0.45$. 
Here, we disregard corrections to $\delta_{c}$ from, e.g., QCD equation of state~\cite{Byrnes:2018clq}. 
The quantity $\tilde{\beta}(m,m_{H})$ is a mass distribution function of \acp{PBH}, which were produced when the horizon mass was $m_{H}$. 
The subscript $_{\mathrm{eq}}$ stands for cosmological quantities at the epoch of matter-radiation equality.  
Cosmic temperature $T$ is related to $m_{H}$ enclosed by the Hubble horizon, i.e.,~\cite{Wang:2019kaf}
\begin{equation}
    \frac{m_{H}}{M_\odot} = 4.76\times10^{-2}  \left(\frac{T}{\mathrm{GeV}}\right)^{-2} \left[\frac{g_{\ast,\rho}(T)}{106.75}\right]^{-\frac{1}{2}} \ ,\label{eq:TmH}
\end{equation}
where $M_\odot$ denotes the solar mass. 
We can relate $f$ with $m_{H}$ by combining Equation~(\ref{eq:fT}) with Equation~(\ref{eq:TmH}) and by reducing $T$ from both equations.
%, or straightforwardly refer to Eq.~(1) in Ref.~\cite{Wang:2019kaf}. 
The coarse-grained fluctuations in the radiation-dominated cosmos are given by~\cite{Young:2014ana,Ando:2018qdb}
\begin{equation}
    \Delta^{2}(k) = \frac{16}{81}\int d\ln q\ \left[w\left(\frac{q}{k}\right)\right]^{2} \left(\frac{q}{k}\right)^{4} \mathcal{T}^{2}\left({q},\frac{1}{k}\right) \mathcal{P}_{\mathcal{R}}(q)\ ,
\end{equation}
where $w(y)=\mathrm{exp}(-y^{2}/2)$ is a Gaussian window function and $\mathcal{T}(q,\tau)=3(\sin{x}-x\cos{x})/x^{3}$ with $x=q\tau/\sqrt{3}$ is a scalar transfer function. 
%We have $\mathcal{T}(q/k)=\mathcal{T}_s(q,\tau)$ with $\mathcal{T}_s$ being a scalar transfer function and $\tau=1/k$.  
We further define the total abundance of \acp{PBH} in dark matter to be $\overbar{f}_{\mathrm{pbh}}=\int f_{\mathrm{pbh}}(m) d \ln m$, and define the average mass of \acp{PBH} to be $\overbar{m}=\int f_{\mathrm{pbh}}(m) dm /\overbar{f}_{\mathrm{pbh}}$.
The latter roughly displays $\overbar{m}\simeq m_\ast$, where $m_\ast$ is corresponded to $k_\ast$.

\begin{figure}[H]

  \hspace{-7pt}   \includegraphics[width=1.0\columnwidth]{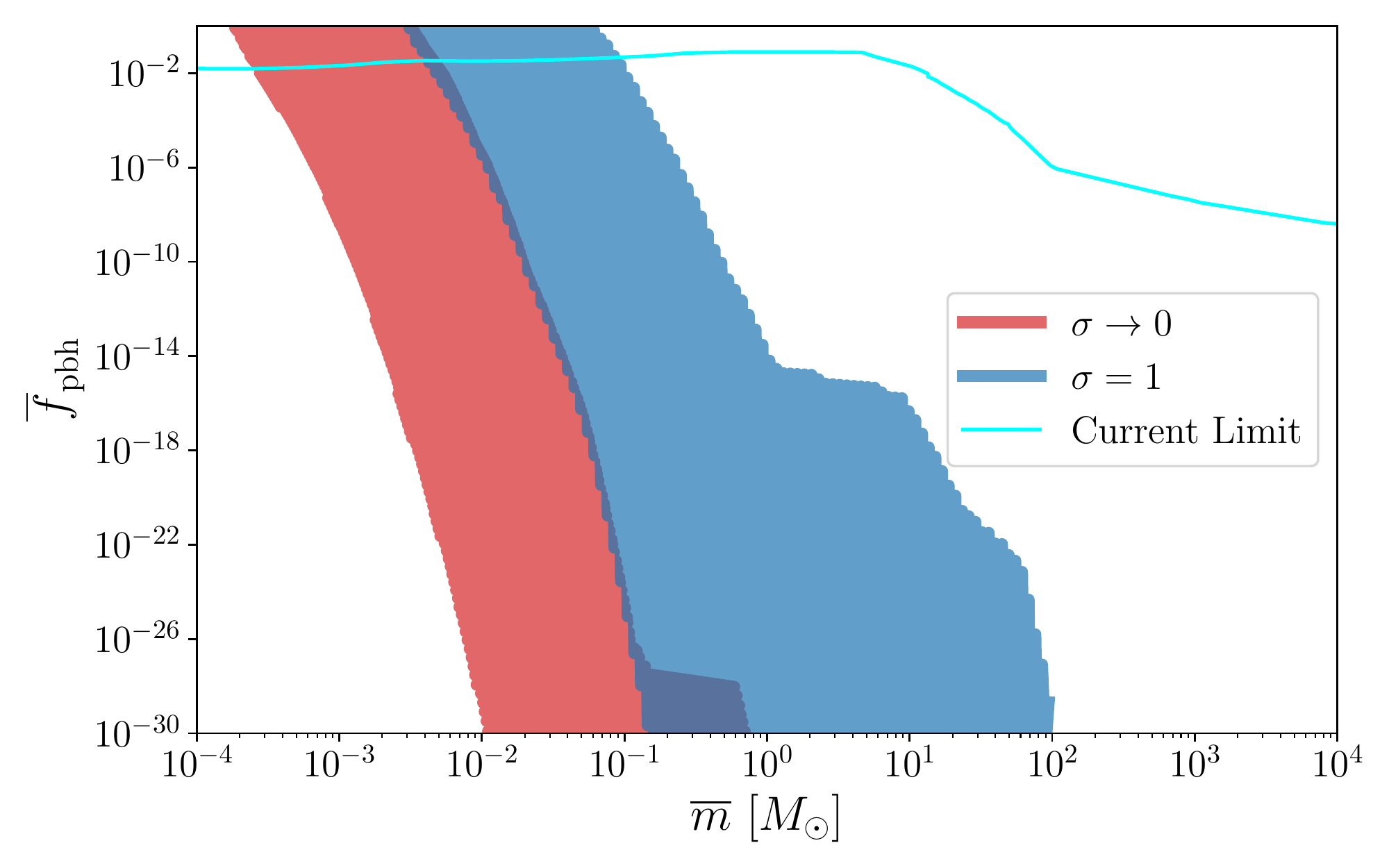}\vspace{-3pt} 
    \caption{The NANOGrav constraints on the averaged mass and total abundance of PBHs. Shaded regions are allowed by the NANOGrav dataset for $\sigma\rightarrow
    0$ and $\sigma=1$. Cyan curve denotes the existing upper limit on the mass function of PBHs~\cite{Carr:2020gox}. }
    \label{fig:fig8}
\end{figure}

Once the constraints on the power spectrum of primordial curvature perturbations are obtained, as shown in Figure~\ref{fig:fig1}, we can recast them into constraints on the mass function of \acp{PBH}, or more precisely, on the average mass and total abundance of \acp{PBH}. 
We show the results in Figure~\ref{fig:fig8}. 
The shaded regions are allowed by the \ac{NANOGrav} 12.5-year dataset for $\sigma\rightarrow0$ (red region) and $\sigma=1$ (blue region). 
When at least one fraction (e.g., $f_{\mathrm{pbh}}=10^{-10}$) of dark matter is composed of \acp{PBH}, the mass range $2\times10^{-4}-10^{-2}M_\odot$ ($4\times10^{-3}-0.2 M_\odot$) for $\sigma\rightarrow0$ ($\sigma=1$) is preferred by the current dataset. 
Based on Figure~3 in Wang et al. ~\cite{Wang:2019kaf}, we find that these mass ranges can be cross-checked with high-precision by observing the \acp{GW} emitted from inspiraling stage of \ac{PBH} binaries. 
In addition, they might be further tested by measuring the anisotropies in \ac{SGWB}~\cite{Wang:2021djr}. 
For comparison, we also depict the existing upper limit (cyan curve) on the mass function of \acp{PBH}, as reviewed in Carr et al. ~\cite{Carr:2020gox}. 
We find the \ac{SIGW} probe to be more powerful than electromagnetic probes, implying that a larger parameter space can be explored with the \ac{SIGW} probe. 
This can also be understood by revisiting Figure~\ref{fig:fig1}, in which we depicted the curves labeling $\overbar{f}_{\mathrm{pbh}}=10^{-10}$. 
The latter is corresponded to $\mathcal{A}\simeq\mathrm{few}\times10^{-2}$. 
In contrast, the \ac{NANOGrav} experiment has reached $\mathcal{A}\simeq10^{-2}$ in the most sensitive frequency band $\sim$$(10^{-9}-10^{-8})$ Hz.

\vspace{-5pt} 

%To be specific, we find the \acp{PBH} to be allowed in the mass range from $10^{-?}M_\odot$ to $10^{-?}M_\odot$ at 95\% confidence level, implying that at least a fraction of dark matter can be accounted for by these \acp{PBH}.  
%Different choices of $\sigma$ do not alter the preferred mass range of \acp{PBH}, but change the abundance of \acp{PBH}, as expected in previous sections. 
%For $\sigma=1$, we find that the abundance of \acp{PBH} is larger than the existing upper limits from other observations. 
%In contrast, for $\sigma\rightarrow0$, we find that the \acp{PBH} are allowed to be a significant component of dark matter. 
%Therefore, the \ac{NANOGrav} dataset supports a relatively narrow distribution function for the power spectrum of primordial curvature perturbations. 

\section{Constraints from Ongoing and Planned Gravitational-Wave Detectors}\label{sec:pros}

% As the mass of primordial black holes remains unknown, we need to explore all possible parameter spaces. Joint observations from detectors such as NanoGrav, SKA, $\mu$Ares, BBO, ET, AdvLIGO, etc., can explore all parameter spaces that may be probed in the future.
%
In the future, the power spectrum of primordial curvature perturbations and the mass function of \acp{PBH}, which remain unknown until now, can be further explored with ongoing and planned \ac{GW} experiments, such as \ac{SKA}~\cite{dewdney2009square,Weltman:2018zrl,Moore:2014lga}, $\mu$Ares~\cite{Sesana:2019vho}, \ac{LISA}~\cite{LISA:2017pwj,Robson:2018ifk}, \ac{BBO}~\cite{Crowder:2005nr,Harry:2006fi}, \ac{DECIGO}~\cite{Sato:2017dkf,Kawamura:2020pcg}, \ac{ET}~\cite{Punturo:2010zz} and Advanced LIGO and Virgo~\cite{Harry:2010zz,VIRGO:2014yos,Somiya:2011np}. 
%Their detailed setups can be found in Ref.~\cite{}. 
Such multi-band observations could explore all possible parameter spaces of primordial curvature perturbations and \acp{PBH}. 
Other experiments are not considered in this work, but our method can be generalized straightforwardly to study them, if needed.

To estimate the sensitivity curve of a given \ac{GW} experiment consisting of $n_{\mathrm{det}}$ detectors, we define an optimal \ac{SNR} denoted with $\rho$ as follows~\cite{Schmitz:2020syl}
\begin{equation}\label{eq:snr}
    \rho^{2} = n_{\mathrm{det}} T_{\mathrm{obs}} \int_{f_{\mathrm{min}}}^{f_{\mathrm{max}}} \left[\frac{S_{\mathrm{GW}}(f)}{S_{n}^{\mathrm{eff}}(f)}\right]^{2} df\ ,
\end{equation}
where $T_{\mathrm{obs}}$ is a duration time of observing run, the \ac{PSD} of \acp{SIGW} is defined as \begin{equation}
    S_{\mathrm{GW}}(f) = \left(\frac{3H_{0}^{2}}{2\pi^{2}f^{3}}\right) \Omega_{\mathrm{GW},0}(f)\ ,
\end{equation}
and the effective noise \ac{PSD} of the detector network is denoted as $S_{n}^{\mathrm{eff}}(f)$ that is a function of $f$. 
The concrete setups of aforementioned experiments are summarized in Table 2 of Ref.~\cite{Campeti:2020xwn} and references therein. 
We consider a single detector for \ac{LISA}, two independent detectors for $\mu$Ares, \ac{BBO} and \ac{DECIGO}, three detectors for \ac{ET}, and 200 pulsars for \ac{SKA}. 
For comparison, we consider one detector for Advanced LIGO with an observing duration of four years and 100\% duty circle.

\begin{figure}[H]

 \hspace{-3pt}    \includegraphics[width=\columnwidth]{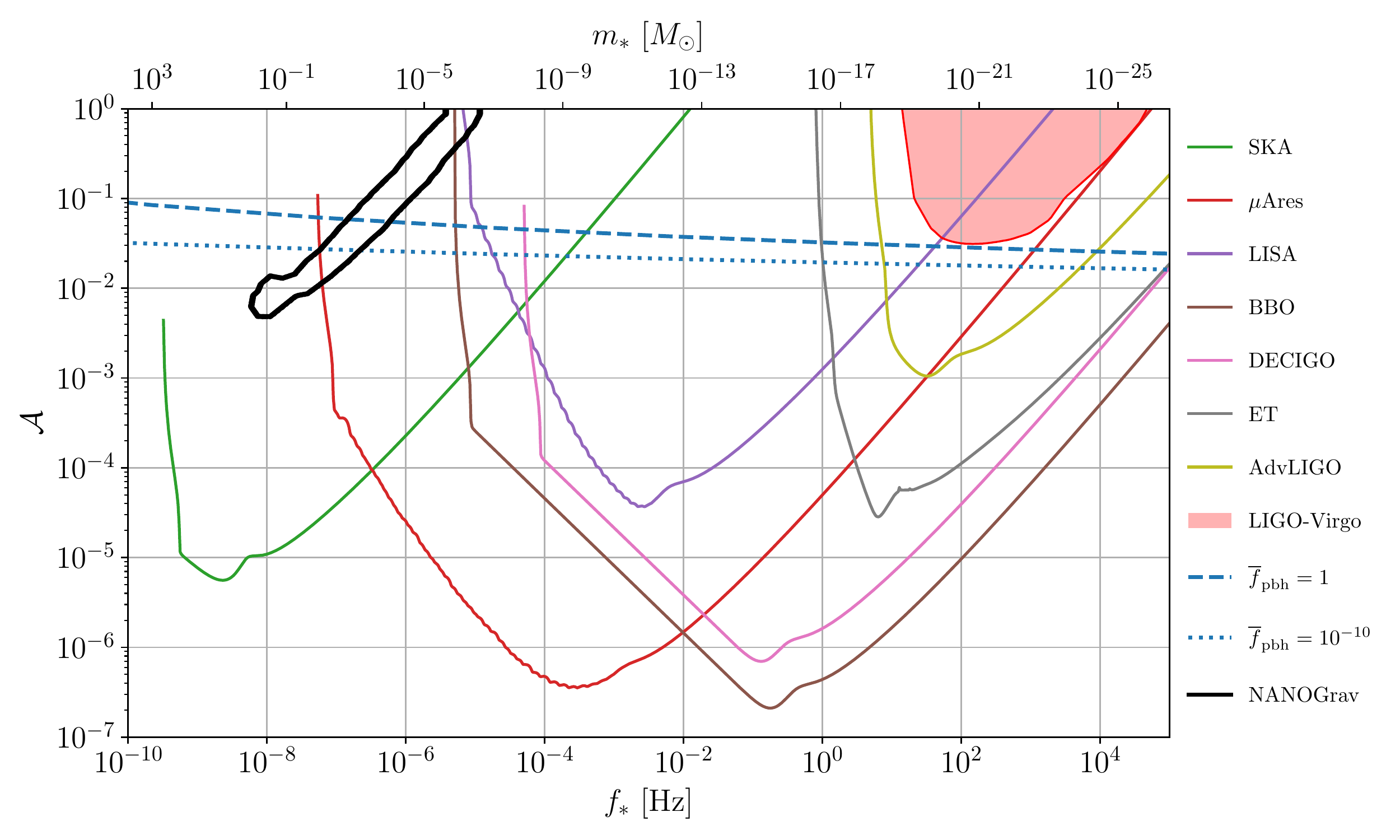}
    \caption{Sensitivities of ongoing and planned GW experiments on measurements of the power spectrum of primordial curvature perturbations with $\sigma\rightarrow0$. The excluded region by Advanced LIGO--Virgo first three observing runs~\cite{Romero-Rodriguez:2021aws} is shown for comparison. We show the allowed region from NANOGrav 12.5-year dataset, as shown in Figure~\ref{fig:fig1}. We also depict critical curves corresponded to $\overbar{f}_{\mathrm{pbh}}=1$ and $\overbar{f}_{\mathrm{pbh}}=10^{-10}$. }
    \label{fig:fig5}
\end{figure}

\begin{figure}[H]

   \hspace{-3pt} \includegraphics[width=\columnwidth]{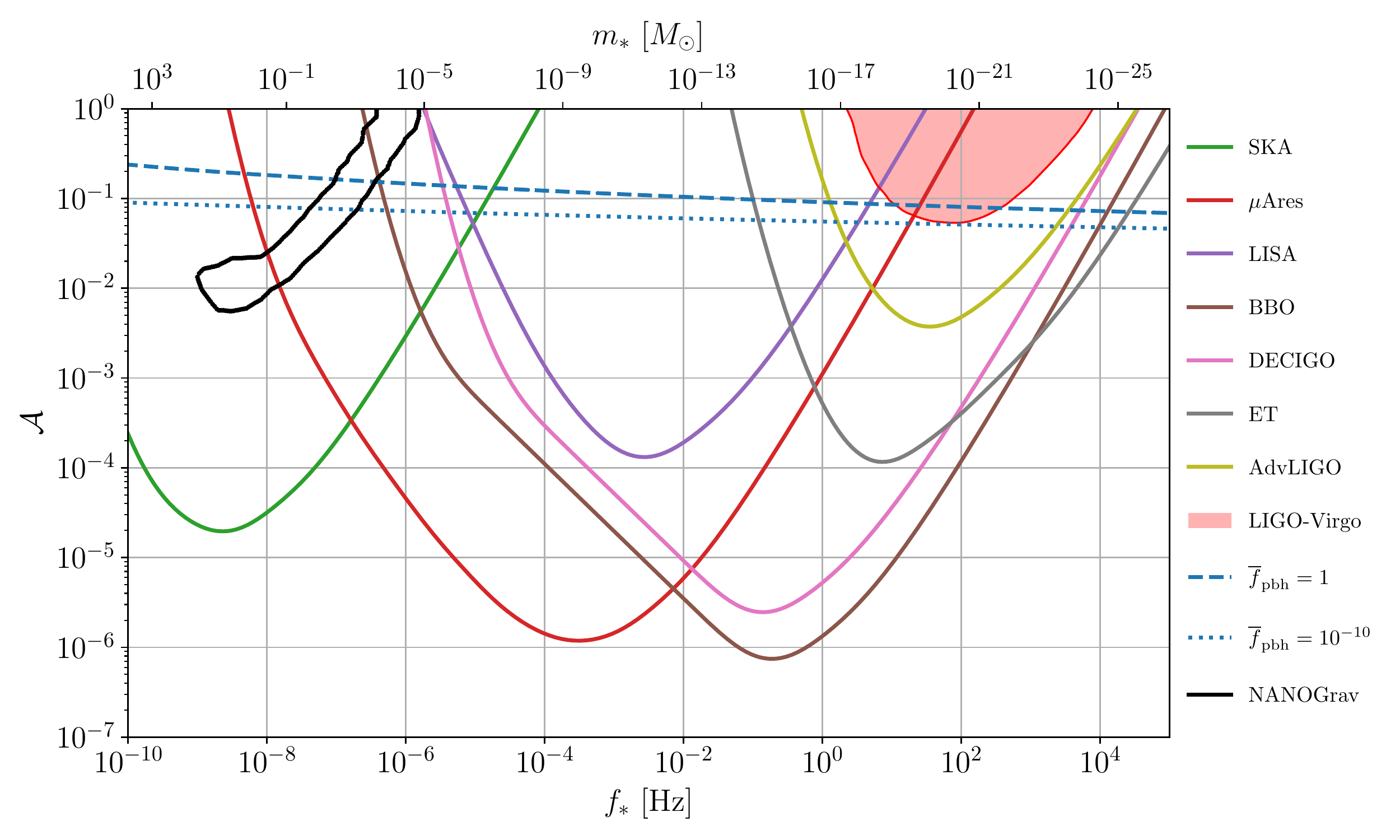}\vspace{-3pt} 
    \caption{The same as Figure~\ref{fig:fig5} but $\sigma=1$. }
    \label{fig:fig9}
\end{figure}

Given a desired value of \ac{SNR}, which is unity in this work, we obtain the minimal detectable $\mathcal{A}_{\mathrm{min}}$ for the given experiment by resolving Equation~(\ref{eq:snr}). 
Since $\Omega_{\mathrm{GW,0}}(f)$ is uniquely determined by $\mathcal{A}$ and $f_{\ast}$, we depict the theoretical expectation of $\mathcal{A}_{\mathrm{min}}$ in the $\mathcal{A}-f_{\ast}$ plane in Figures~\ref{fig:fig5} and \ref{fig:fig9} for the choices of $\sigma\rightarrow0$ and $\sigma=1$, respectively. 
For comparison, we plot the exclusion region on $\mathcal{A}$ from 
%the \ac{CMB} and \ac{BBN} (purple shaded)~\cite{Giovanetti:2021izc} and 
the Advanced LIGO and Virgo first three observing runs (red shaded)~\cite{Romero-Rodriguez:2021aws}. 
%To get the former, we resolve $\int \Omega_{\mathrm{GW}}(f)d\ln f \leq\Delta N_{\mathrm{eff}}$ via taking into account a 95\% CL upper limit on the effective relativistic degrees of freedom $\Delta N_{\mathrm{eff}}$~\cite{Planck:2018vyg}. 
%Here, the solid curves stand for $\sigma\rightarrow0$ while the dashed ones stand for $\sigma=1$. 
\vspace{-3pt} 

Based on Figures~\ref{fig:fig5} and \ref{fig:fig9}, we find that the regions allowed by the current dataset of \ac{NANOGrav} (enclosed by black curves) can be thoroughly tested with \ac{SKA} and $\mu$Ares. 
Smaller values of $\mathcal{A}$, i.e., $\sim$$(10^{-4}-10^{-7})$, are also expected to be reached by the planned \ac{GW} experiments. 
The sensitivity of Advanced LIGO to measure the \acp{SIGW} is expected to be improved by one order of magnitude in upcoming observing runs. 
If other values of \ac{SNR} are desired, we can obtain revised $\mathcal{A}_{\mathrm{min}}$ by rescaling the above results by multiplying a factor of $(\mathrm{SNR})^{1/2}$. 
After recasting the expected constraints on $\mathcal{A}$ into constraints on the mass function of \acp{PBH}, we further find that the full parameter space of \acp{PBH}, that account for at least a fraction of dark matter, e.g., $f_{\mathrm{pbh}}\sim10^{-10}$, can be thoroughly explored with multi-band \ac{GW} measurements, e.g., a detector network composed of \ac{SKA}, $\mu$Ares and \ac{ET}, or other detector networks. 
Therefore, we expect that the scenario of \acp{PBH} as a candidate of dark matter can be supported or vetoed by future \ac{GW} observations.

\section{Conclusions and Discussion}\label{sec:conc}

\textls[-7]{In this work, we obtained new constraints on the power spectrum of primordial curvature perturbations and the mass function of \acp{PBH} by searching for the energy density fraction spectrum of \acp{SIGW} in the \ac{NANOGrav} 12.5-year dataset. 
We found the lower limits on $\mathcal{A}$, namely, $\mathcal{A}\gtrsim10^{-2}$ and showed the 95\% CL contours of $\mathcal{A}$ and $f_{\ast}$ (see Figure~\ref{fig:fig1}). 
Recasting these contours into the \ac{PBH} mass-abundance plane, we showed the parameter space of \acp{PBH} allowed by the \ac{NANOGrav} (see Figure~\ref{fig:fig8}). 
We found that at least a fraction of dark matter can be interpreted with the scenario of \acp{PBH} in the mass range $\sim$$(10^{-4}-10^{-1})M_{\odot}$.} We also studied dependence of the above results on the value of $\sigma$. 
Furthermore, this mass range is expected to be cross-checked with high-precision by observing the \acp{GW} from \ac{PBH} mergers (see Figure~3 in Wang et al. ~\cite{Wang:2019kaf}). 
We also found that the \ac{SIGW} probe is much more powerful than other probes to search for the \acp{PBH}.

We further forecasted the sensitivity curves of ongoing and planned \ac{GW} experiments on detection of \acp{PBH} and the primordial curvature perturbations by searching for \acp{SIGW}. 
We found that the primordial curvature perturbations with spectral amplitude larger than $\sim$$(10^{-4}-10^{-7})$ can be measured with planned \ac{GW} detection programs. 
The sensitivity of Advanced LIGO-Virgo to detect the \acp{SIGW} was expected to be improved by one order of magnitude in the near future. 
This prediction may promote extensive investigations of cosmic inflation at late-time stages. 
Meanwhile, the scenario of \acp{PBH} within almost whole mass range can be thoroughly tested, since the critical spectral amplitude to form the \acp{PBH} is much larger than $10^{-4}$ (see Figures~\ref{fig:fig5} and \ref{fig:fig9}). 
In particular, we can search for the \acp{PBH} within mass range $(10^{-16}-10^{-11})M_\odot$, which can compose all of the dark matter and otherwise are beyond capabilities of other probes. 
In addition, we emphasized the importance of multi-band \ac{GW} detector networks for accomplishing the above theoretical expectations.

In this paper, we made several assumptions to simplify our computations. 
First, we disregarded possible contributions of primordial non-Gaussianity to the formation of \acp{PBH}~\cite{Franciolini:2018vbk,Gow:2022jfb,Cai:2022erk,Ferrante:2022mui,Kitajima:2021fpq} and to the production of \acp{SIGW}~\cite{Adshead:2021hnm}. It is an interesting topic to study the primordial non-Gaussianity, deserving an independent work.  
Second, we took into account the median value of the effective relativistic degrees of freedom of the early Universe, but disregarded their uncertainties~\cite{Saikawa:2018rcs}. 
In fact, changing the above two assumptions could alter the mass function of \acp{PBH} and the energy density fraction spectrum of \acp{SIGW}. 
We would leave a possible study of this question to future works. 
Third, we disregarded contributions of nonlinear cosmological perturbations to the energy density fraction spectrum of \acp{SIGW}, since there is not a complete theory of \acp{SIGW} at the third order~\cite{Yuan:2019udt,Zhou:2021vcw,Chang:2022nzu,Chen:2022dah,Meng:2022ixx}. 
We might revisit this assumption in future, once the theory is complete.

\authorcontributions{{Conceptualization, S.W.; methodology, Z.C.Z. and S.W.; software, Z.C.Z.; validation, Z.C.Z. and S.W.; formal analysis, Z.C.Z. and S.W.; investigation, Z.C.Z. and S.W.; resources, S.W.; data curation, Z.C.Z.; writing---original draft preparation, S.W.; writing---review and editing, Z.C.Z. and S.W.; visualization, Z.C.Z. and S.W.; supervision, S.W.; project administration, S.W.; funding acquisition, Z.C.Z. and S.W.. All authors have read and agreed to the published version of the manuscript.} 
}

\funding{{This research was funded by the National Natural Science Foundation of China grant number 12175243 and grant number 12005016, the Key Research Program of the Chinese Academy of Sciences grant number XDPB15, and the science research grants from the China Manned Space Project grant number CMS-CSST-2021-B01.} 
}

\dataavailability{{Publicly available datasets were analyzed in this study. This data can be found here: https://data.nanograv.org} 
}

\acknowledgments{
We acknowledge {Zu-Cheng Chen,} 
 Jun-Peng Li and Ke Wang for helpful discussions.}

\conflictsofinterest{{The authors declare no conflict of interest.} }

\begin{adjustwidth}{-\extralength}{0cm}

\printendnotes[custom] % Un-comment to print a list of endnotes

\reftitle{References}

% \bibliographystyle{h-physrev}

%=====================================
% References, variant B: internal bibliography
%=====================================
\PublishersNote{}
\end{adjustwidth}
\end{document}